\documentclass[twocolumn]{svjour3}         % twocolumn
\smartqed  % flush right qed marks, e.g. at end of proof
\usepackage{graphicx}
\begin{document}

\title{Studying the first galaxies with ALMA } \subtitle{}


\author{C.L. Carilli, F. Walter, R. Wang, A. Wootten, K. Menten, F. Bertoldi, 
E. Schinnerer, P. Cox, A. Beelen, A. Omont}


\institute{C.~Carilli  National Radio Astronomy Observatory, Socorro\\
              \email{ccarilli@nrao.edu} \\
           F.~Walter   Max Planck Insitut f\"ur Astronomie, Heidelberg\\
           R. Wang  National Radio Astronomy Observatory, \\
           A. Wootten  National Radio Astronomy Observatory, \\
           K. Menten  Max-Planck Institute for Radio Astronomy, Bonn\\
           F. Bertoldi  Bonn University, Bonn\\
           E. Schinnerer  Max Planck Insitut f\"ur Astronomie Heidelberg,\\
           P. Cox  IRAM, Grenoble \\
           A. Beelen  Max-Planck Institute for Radio Astronomy,\\
           A. Omont    Institute de Astrophysique, Paris \\
}

\date{Received: date / Accepted: date}

\maketitle

\begin{abstract}

We discuss observations of the first galaxies, within cosmic
reionization, at centimeter and millimeter wavelengths. We present a
summary of current observations of the host galaxies of the most
distant QSOs ($z \sim 6$). These observations reveal the gas, dust,
and star formation in the host galaxies on kpc-scales. These data
imply an enriched ISM in the QSO host galaxies within 1 Gyr of the big
bang, and are consistent with models of coeval supermassive black hole
and spheroidal galaxy formation in major mergers at high
redshift. Current instruments are limited to studying truly pathologic
objects at these redshifts, meaning hyper-luminous infrared galaxies
($L_{FIR} \sim 10^{13}$ L$_\odot$). ALMA will provide the one to two
orders of magnitude improvement in millimeter astronomy required to
study normal star forming galaxies (ie. Ly-$\alpha$ emitters) at $z
\sim 6$.  ALMA will reveal, at sub-kpc spatial resolution, the thermal
gas and dust -- the fundamental fuel for star formation -- in galaxies
into cosmic reionization.

% \keywords{First keyword \and Second keyword \and More}
\end{abstract}

\section{Introduction: Cosmic reionization and the first galaxies}
\vspace*{-2.5mm}

Observations of the first generation of galaxies and supermassive
black holes (SMBH) provide the greatest leverage into theories of
cosmic structure formation, and are a principle science driver for
all future large area telescopes, from meter to Xray wavelengths.  The
recent discovery of the Gunn-Peterson effect, ie. Ly-$\alpha$
absorption by a partially neutral intergalactic medium (IGM), toward
the most distant ($z \sim 6$) QSOs indicates that we have finally
probed into the near-edge of cosmic reionization (Fan et al. 2006).
Reionization sets a fundamental benchmark in cosmic structure
formation, indicating the formation of the first luminous objects
which act to reionize the IGM.  Detection of large scale polarization
of the CMB, corresponding to Thomson scattering of the CMB by the IGM
during reionization, suggests a significant ionization fraction 
extending to $z \sim 11 \pm3$ (Page et al. 2006). Overall, current
data indicate that cosmic reionization is a complex process, with
significant variance in space and time (Fan, Carilli,
Keating 2006). The on-set of Gunn-Peterson absorption at $z \ge 6$
implies that the IGM becomes opaque at observed wavelengths $\le
1\mu$m, such that observations of the first luminous objects will
be limited to radio through near IR wavelengths. 

In this contribution we will discuss the current status of centimeter
and millimeter observations of the most distant objects. We will then
discuss the revolution afforded by ALMA in this area of research,
taking the field from the current study of truly pathologic, rare
objects, to the study of the first generation of normal star forming
galaxies. In this contribution we concentrate on molecular line and
dust continuum emission. Walter \& Carilli (this volume) discuss the
exciting prospects for studying the fine structure PDR cooling lines,
such as [CII], from the first galaxies.

\section{Current centimeter and millimeter observations of
$z \sim 6$ objects}
\vspace*{-2.5mm}

\subsection{The host galaxies of $z \sim 6$ SDSS QSOs}

Given the sensitivities of current instruments, observations of
sources at $z \sim 6$ are restricted to Hyper Luminous Infrared
galaxies (ie. $L_{FIR} > 10^{13}$L$_\odot$). At these extreme
redshifts, the samples remain limited to the host galaxies of
optically luminous QSOs selected from the SDSS. The study of such
systems has become paramount since the discovery of the bulge mass --
black hole mass correlation in nearby galaxies, a result which
suggests a fundamental relationship between black hole and spheroidal
galaxy formation (Gebhardt et al. 2000).  Our millimeter surveys of
the $z \sim 6$ QSOs show that roughly 1/3 of optically selected QSOs
are also hyper-luminous infrared galaxies ($L_{FIR} \ge 10^{13}$
L$_\odot$), emitting copious thermal radiation from warm dust (Figure
1; Wang et al. 2007). This corresponds to roughly 10$\%$ of the
bolometric luminosity of the QSO (which is dominated by the AGN 'big
blue bump'), and the question remains open as to the dominant dust
heating mechanism: star formation or the AGN?

\begin{figure}
\begin{center}
  \includegraphics[scale=0.5]{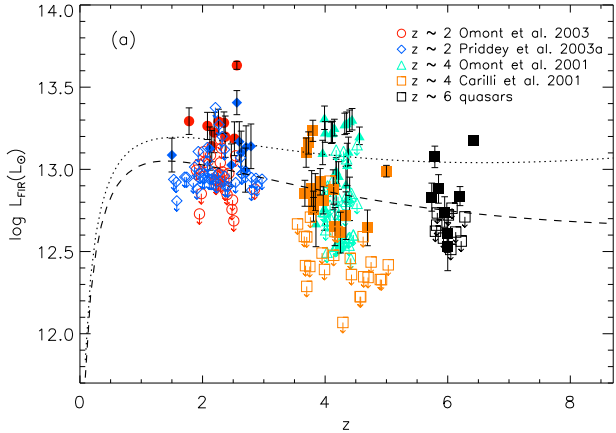}
\end{center}
\caption{The logarithm of the FIR luminosity versus redshift for
different QSO samples observed at (sub)mm wavelengths (Wang et al. 2007).
The open symbols with arrows denote upper limits.
The dashed and dotted lines represent the typical 3$\sigma$
detection limits of MAMBO at 250 GHz and SCUBA at 350GHz, namely 
$S_{250} = 2.4$mJy, and $S_{350} = 8.5$  mJy, respectively. 
}
\label{fig:1}       % Give a unique label
\end{figure}

The best studied of the $z \sim 6$ QSOs is the most distant QSO known,
J1148+5251, at $z = 6.419$.  This galaxy been detected in thermal
dust, non-thermal radio continuum, and CO line emission (Walter et
al. 2003; Bertoldi al. 2003; Carilli et al. 2004), with an implied
dust mass of $7\times 10^8$ M$_\odot$, and a molecular gas mass of
$2\times 10^{10}$ M$_\odot$.  The molecular gas is extended over
$\sim 1"$, or $\sim 5.5$ kpc (Figure 2). High resolution VLA imaging
of the molecular gas distribution in J1148+5251 provides the only
direct measure of the host galaxy dynamical mass, resulting in a value
of $\sim 4\times10^{10}$ M$_\odot$ within 3 kpc of the galaxy center
(Walter et al. 2004; Walter \& Carilli, this volume).  This mass is
comparable to the gas mass, suggesting a baryon-dominated potential
for the inner few kpc of the galaxy (Lintott et al. 2006), as is true
in nearby spheroidal galaxies, and for ULIRGs (Downes \& Solomon
1998).  The dynamical mass is also more than an order of magnitude
lower than expected based on the bulge mass -- black hole mass
correlation, suggesting a departure from this fundamental relationship
at the highest redshifts, with the SMBH forming prior to the
spheroidal galaxy (Walter et al.  2004).

\begin{figure}
\begin{center}
  \includegraphics[scale=0.35]{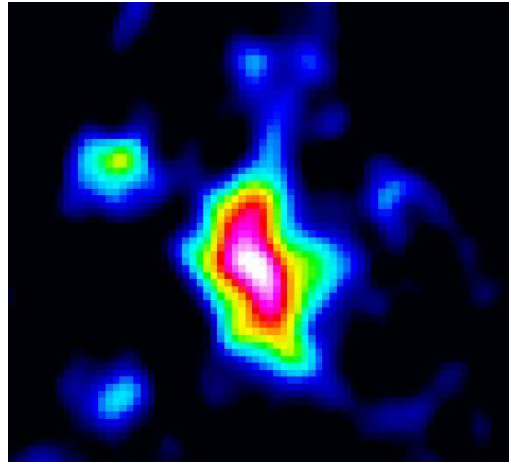}
\end{center}
\caption{The velocity integrated 
CO 3-2 emission from J1148+5251 imaged by the 
VLA at 0.4" resolution (Walter et al. 2004). 
The total flux  is 0.15 Jy km s$^{-1}$, and
the source is clearly resolved, with a full extend of about 
1$"$. The figure is about 3$"$ on a side.}
\label{fig:1}       % Give a unique label
\end{figure}

\begin{figure}
\begin{center}
  \includegraphics[scale=0.5]{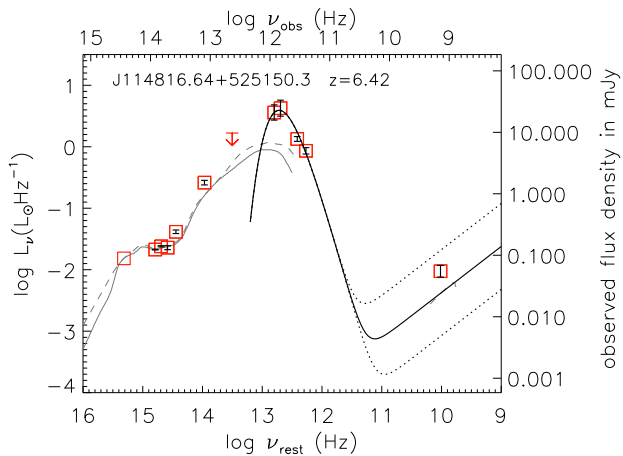}
\end{center}
\caption{The SED of J1148+5251 from the rest frame near-IR to
the radio (Jiang et al. 2007; Wang et al. 
2007; Beelen et al. 2006). The models at rest frame frequencies, $\nu > 10^{13}$ Hz 
are two standard QSO SEDs including emission from hot ($\sim 1000$K) dust. 
The rest-frame far-IR through radio model entails a 55K modified black body,
plus synchrotron radio emission that follows the radio-FIR 
correlation for star forming galaxies (Beelen et al. 2006; Wang
et al. 2007). The dashed lines indicate the range defined by
star forming galaxies (Yun et al. 2000). 
}
\label{fig:1}       % Give a unique label
\end{figure}

The radio through near-IR SED of J1148+5251 is shown in Figure 3
(Beelen et al. 2005). The Spitzer bands are consistent with the
standard QSO optical through mid-IR SED, including a hot dust
component ($\sim 1000$K), presumably heated by the AGN. However, the
observed (sub)mm data reveal a clear rest-frame FIR excess. The rest
frame FIR through radio SED is reasonably fit by a template that
follows the radio through FIR correlation for star forming galaxies
(Yun et al. 2000), with a dust temperature of 55K.  The implied star
formation rate is of order 3000 M$_\odot$ year$^{-1}$ (Bertoldi et
al. 2003).  Most recently, IRAM 30m observations of J1148+5251 have
yielded the first detection of the fine structure line of [CII] at
cosmologically significant redshifts (Maiolino et al. 2005). This line
is thought to dominate ISM cooling in photon-dominated regions,
ie. the interface regions between giant molecular clouds and HII
regions.

These observations of J1148+5251 demonstrate that large reservoirs of
dust and metal enriched atomic and molecular gas can exist in the most
distant galaxies, within 870 Myr of the big bang.  The molecular gas
and [CII] emission suggest a substantially enriched ISM on kpc-scales.
The molecular gas represents the requisite fuel for star formation.

The mere existence of such a large dust mass so early in the universe
raises the interesting question: how does a galaxy form so much dust
so early in the universe?  The standard mechanism of dust formation in
the cool winds from low mass (AGB) stars takes a factor two or so too
long. Maiolino et al. (2004) and Strata et al. (2007) suggest dust
formation associated with massive stars in these distant galaxies.
They show that the reddening toward the most distant objects is
consistent with different dust properties (ie. silicates and amorphous
carbon grains), as expected for dust formed in type-II SNe (although
cf. Venkatesan et al. 2006).

Overall, we conclude that J1148+5251 is a likely candidate for the
co-eval formation of a SMBH through Eddington-limited accretion, and a
large spheroidal galaxy in a spectacular starburst, within 1 Gyr of
the big bang. This conclusion is consistent with the general notion of
'downsizing' in both galaxy and supermassive black hole formation
(Cowie et al. 1996; Heckman et al. 2004), meaning that the most
massive black holes ($>10^9$ M$_\odot$) and galaxies ($> 10^{12}$
M$_\odot$) may form at high redshift in extreme, gas rich
mergers/accretion events.

Li et al. (200) and Robertson et al. (2007) have performed detailed
modeling of a system like J1148+5251, including feedback from the AGN
to regulate star formation. They show that is plausible to form both
the galaxy and the SMBH in rare peaks in the cosmic density field
(comoving density $\sim 10^{-9}$ Mpc$^{-3}$), through a series of
major mergers of gas rich galaxies, starting at $z \sim 14$, resulting
in a SMBH of $\sim 10^9$ M$_\odot$, and a galaxy of total stellar mass
$\sim 10^{12}$ M$_\odot$ by $z \sim 6$.  The system will eventually
evolve into a rare, extreme mass cluster ($\sim 10^{15}$ M$_\odot$)
today. The ISM abundance in the inner few kpc will quickly rise to
$\sim$ solar, although the dust formation mechanism remains uncertain.

In support of this conclusion, Figure 4 shows the relationship between
FIR luminosity and bolometric luminosity for the (sub)mm detected $z
\sim$ QSOs, as well as a low redshift sample of optically selected
QSOs (eg. PG sample), and an IRAS selected sample of QSOs (Wang et
al. 2007; Hao et al. 2005). The $z \sim 6$ sources fall at the extreme
luminosity end of the sample, but interestingly, they also follow the
trend in FIR to bolometric luminosity set by the low redshift IRAS
QSOs, and as opposed to the trend set by the optically selected
QSOs. The optically selected QSOs typically have early-type host
galaxies, while the IRAS selected QSOs reside in major mergers, with
co-eval starbursts (Hao et al.  2005).

\begin{figure}
\begin{center}
  \includegraphics[scale=0.5]{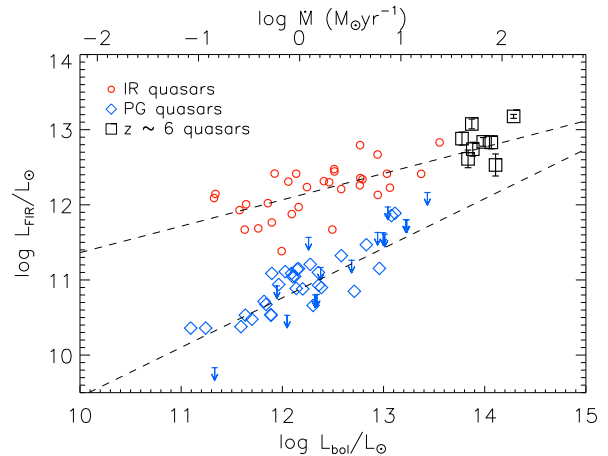}
\end{center}
\caption{Correlation of $L_{FIR}$--$L_{bol}$ for QSO host galaxies. The (sub)mm
detected z$\sim$6 quasars are plotted as black squares with error
bars denoting 1$\sigma$ r.m.s.. The local IR and PG quasars from Hao
et al. (2005) are plotted as red circles and blue diamonds with
arrows denoting upper limits in $L_{FIR}$. The dashed lines
represent the linear regression results for the two local quasar
samples.}
\label{fig:1}       % Give a unique label
\end{figure}

A final interesting aspect of the molecular line studies of the most
distant QSO host galaxies is the derivation of the sizes of the cosmic
Stromgren spheres (Walter et al. 2003; Fan et al. 2006).  The size of
the ionized region around the QSO, presumably formed by the radiation
from the QSO, can be derived from the difference between the redshift
of the host galaxy and the redshift of the on-set of the Gunn-Peterson
trough.  For J1148+5251, this redshift difference is $\Delta z \sim
0.1$, implying a physical radius for the cosmic Stromgren sphere of $R
= 4.7$Mpc. This radius can be related to the cosmic neutral fraction
using the QSO ionizing luminosity, and the mean baryon density,
through the equation: $t_{q} = 10^5 R^3 f(HI)$, where $t_q$ is the qso
lifetime, and $f(HI)$ is the IGM neutral fraction (White et
al. 2005). For J1148+5251, the implied QSO lifetime is $\sim 10^7
f(HI)$ years. A number of authors (Wyithe et al. 2005; Fan et
al. 2006; Kurk et al. 2007) have inverted this equation in order to
derive the IGM neutral fraction. Using the J1148+5251 CO host
galaxy redshift, plus the redshifts for other $z \sim 6$ QSO host
galaxies derived from low ionization broad lines (eg. MgII), they
derive a mean neutral fraction at $z \sim 6.2$ of $f(HI) > 0.1$,
assuming a fiducial QSO lifetime $\ge 10^6$ years.

\subsection{Limits on normal galaxies: the Cosmos field}

J1148+5251 is an extremely rare and pathologically luminous object,
unlike anything seen nearby. For instance, there are only some 50 or
so of these SDSS $z \sim 6$ QSOs on the entire sky!

We have recently investigated the properties of more normal star
forming galaxies at $z \sim 6$ using the Ly-$\alpha$ emitting galaxies
(LAEs) selected through a wide field, narrow band search of the Cosmos
field (Murayama et al. 2007). The sensitivity to the Ly-$\alpha$ line
is such that one can detect galaxies with star formation rates of
$\sim 10$ M$_\odot$ year$^{-1}$  into cosmic
reionization. These galaxies are numerous, with roughly 100 deg$^{-2}$
in a narrow redshift search range of $z = 5.7 \pm 0.05$.
Extrapolation of the luminosity function to dwarf star forming
galaxies could provide enough photons to reionization the universe
(Fan et al. 2006).

We have taken the sample of $\sim 100$ LAEs from the Cosmos field and
searched for radio and millimeter emission using MAMBO (Bertoldi et
al. 2007) and the VLA (Schinnerer et al. 2007).  We do not detect any
individual source down to 3$\sigma$ limits of $\sim 30\mu$Jy
beam$^{-1}$ at 1.4 GHz, nor do we detect a source in a stacking
analysis, to a 2$\sigma$ limit of $2.5\mu$Jy beam$^{-1}$ (Carilli et
al. 2007).  At 250 GHz we do not detect any of the 10 LAEs that are
located within the central regions of the COSMOS field covered by
MAMBO ($20' \times 20'$) to a typical 2$\sigma$ limit of $S_{250} <
2$mJy.  The radio data imply that there are no low luminosity radio
AGN with $L_{1.4} > 6\times 10^{24}$ W Hz$^{-1}$ in the LAE sample.

These radio and millimeter observations rule out any highly obscured,
extreme starbursts in the sample, ie. any galaxies with massive star
formation rates $> 1500$ M$_\odot$ year$^{-1}$ in the full sample
(based on the radio data), or 500 M$_\odot$ year$^{-1}$ for the 10\%
of the LAE sample that fall in the central MAMBO field. The stacking
analysis implies an upper limit to the mean massive star formation
rate of $\sim 100$ M$_\odot$ year$^{-1}$.

While this study represents the most sensitive, widest field radio and
mm study of $z \sim 6$ LAEs to date, it also accentuates the
relatively poor limits that can be reached in the radio and mm for
star forming galaxies at the highest redshifts, when compared to
studies using the Ly$\alpha$ line.

\section{The ALMA revolution}
\vspace*{-2.5mm}

ALMA will be able to detect the thermal emission from warm dust from a
source like J1148+5251 in 1 second. Moreover, it will detect the more
normal galaxy population ($S_{250} \sim 20\mu$Jy for SFR $\sim 10$
M$_\odot$ year$^{-1}$), in a few hours.

\begin{figure}
\begin{center}
  \includegraphics[scale=0.3,angle=-90]{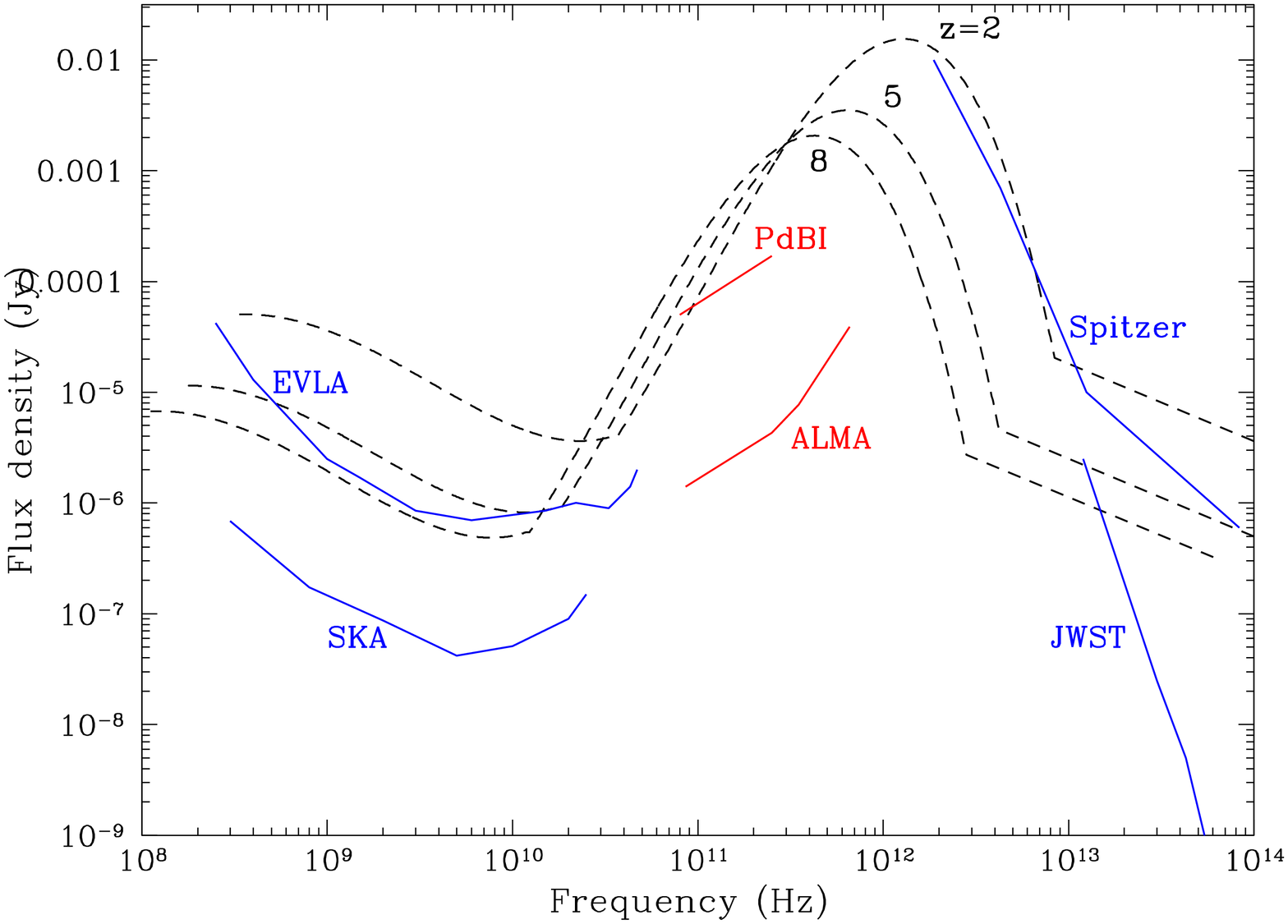}
  \includegraphics[scale=0.3,angle=-90]{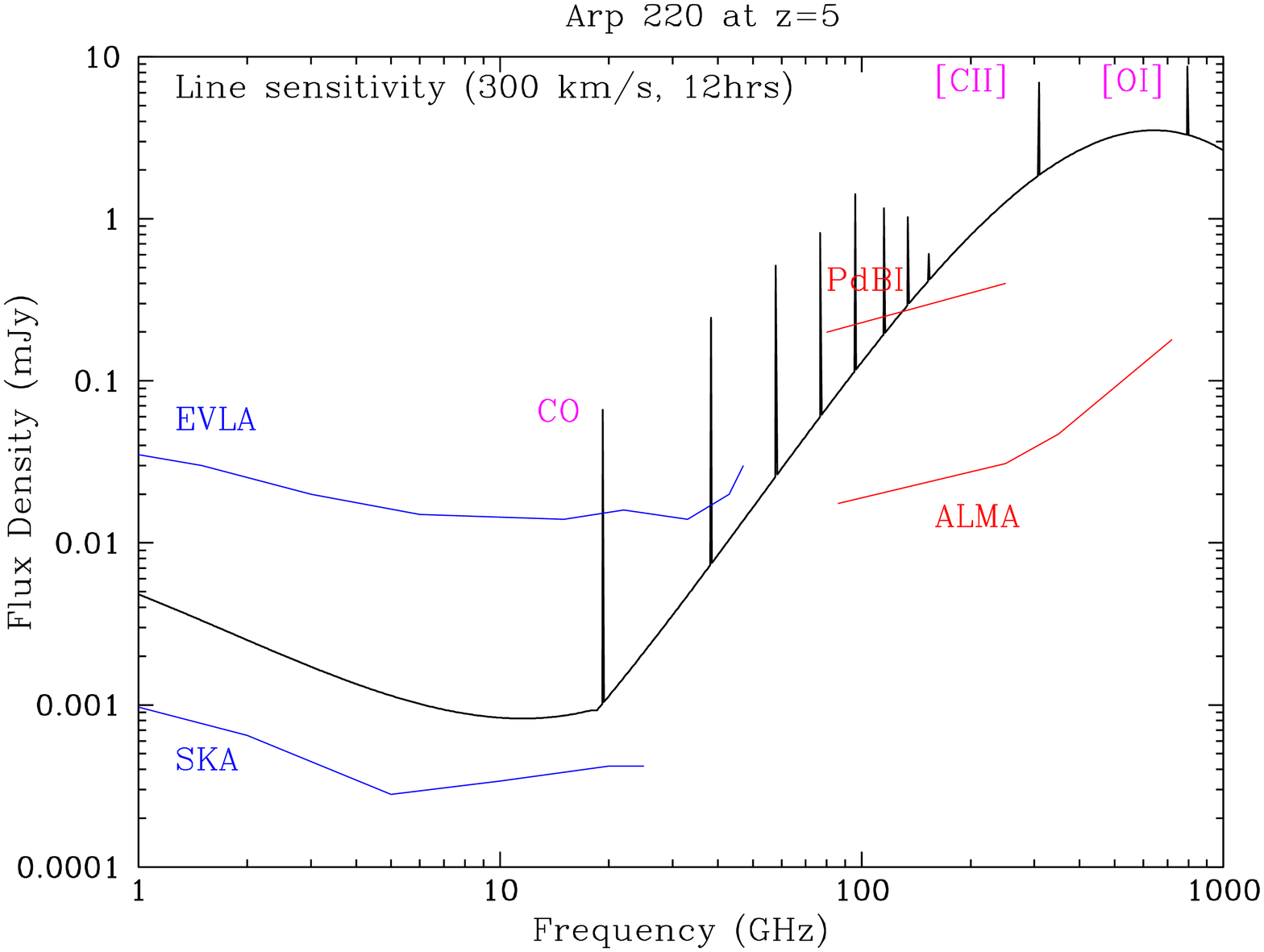}
\end{center}
\caption{Top: Continuum spectrum of an active
star forming galaxy with a star formation
rate $\sim 100$ M$_\odot$ year$^{-1}$,
at $z = 2, 5$, and 8. The curves show the continuum 
sensitivities of various telescopes in 12hours. 
Bottom: Line spectrum of the same galaxy, but only at $z =5$.
The line sensitivities were derived assuming a line width of
300 km s$^{-1}$.
}
\label{fig:3}       % Give a unique label
\end{figure}

\begin{figure}
\begin{center}
  \includegraphics[scale=0.4]{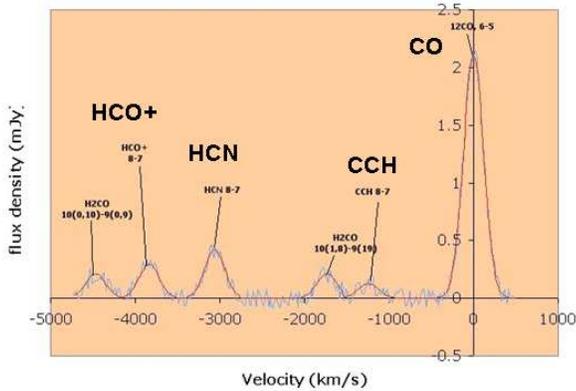}
\end{center}
\caption{A simulation of an ALMA spectrum of J1148+5251 at 
$z  = 6.42$. The spectrum shows the 8GHz bandpass, with an
integration time of 24 hours, centered around 93GHz.
}
\label{fig:3}       % Give a unique label
\end{figure}

Figure 5 shows the sensitivity of current and future telescopes from
the radio through the near-IR, along with the spectrum of an active
star forming galaxy, like Arp 220 (SFR $\sim 100$ M$_\odot$
year$^{-1}$ or $L_{FIR} \sim 10^{12}$ L$_\odot$). The EVLA, and
eventually the SKA, will study the non-thermal (and possibly free-free
thermal) emission associated with star formation, and possibly AGN,
from these distant galaxies, as well as the low order transitions from
molecular gas. The JWST will observe the stars, the AGN, and the
ionized gas. ALMA reveals the thermal emission from dust and gas,
including high order molecular line transitions, and fine structure
ISM cooling lines (Walter \& Carilli, this volume), from the first
galaxies -- the basic fuel for galaxy formation. ALMA provides the
more than an order of magnitude increase in sensitivity and resolution
to both detect, and image at sub-kpc resolution, the gas and dust in
normal star forming galaxies (eg. Ly-$\alpha$ galaxies, with star
formation rates $\sim 10$ M$_\odot$ year$^{-1}$) back to the first
generation of galaxies during cosmic reionization.

As an example, Figure 6 shows the calculated spectrum for J1148+5251
for the 90GHz band of ALMA in 24hours. The CO lines will be detected
with essentially infinite signal-to-noise, allowing detailed imaging
and dynamical studies on sub-kpc scales. Moreover, in a given 8GHz
bandwidth for ALMA, we will detect transitions from numerous
astrochemically interesting molecules, such as the dense, pre-star
forming gas tracers, HCN and HCO+ (Gao et al. 2007).

Galaxy formation is a complex process, and proper studies require a
panchromatic approach. ALMA represents the more than an order of
magnitude increase in sensitivity required to probe normal galaxies
into cosmic reionization.

\noindent We acknowledge support from the Max-Planck Society and the
Alexander von Humboldt Foundation through the Max-Planck
Forshungspreise 2005.  The National Radio Astronomy Observatory is a
facility of the National Science Foundation, operated by Associated
Universities, Inc.

\end{document}